\documentclass[aps,prl,twocolumn,groupedaddress,floats,showpacs]{revtex4}

\usepackage{amssymb}
\usepackage{graphicx}
\usepackage{amsmath}
\usepackage{times}
\usepackage{color}
\usepackage{subfigure}
\usepackage{setspace}
\usepackage{bm}% bold math

\begin{document}

\newcommand {\beq} {\begin{equation}}
\newcommand {\eeq} {\end{equation}}
\newcommand {\bqa} {\begin{eqnarray}}
\newcommand {\eqa} {\end{eqnarray}}
\newcommand {\ba} {\ensuremath{b^\dagger}}
\newcommand {\da} {\ensuremath{d^\dagger}}
\newcommand {\ha} {\ensuremath{h^\dagger}}
\newcommand {\adag} {\ensuremath{a^\dagger}}
\newcommand {\no} {\nonumber}
\newcommand {\ep} {\ensuremath{\epsilon}}
\newcommand {\ca} {\ensuremath{c^\dagger}}
\newcommand {\ga} {\ensuremath{\gamma^\dagger}}
\newcommand {\gm} {\ensuremath{\gamma}}
\newcommand {\up} {\ensuremath{\uparrow}}
\newcommand {\dn} {\ensuremath{\downarrow}}
\newcommand {\ms} {\medskip}
\newcommand {\bs} {\bigskip}
\newcommand{\kk} {\ensuremath{{\bf k}}}
\newcommand{\rr} {\ensuremath{{\bf r}}}
\newcommand{\kp} {\ensuremath{{\bf k'}}}
\newcommand {\qq} {\ensuremath{{\bf q}}}
\newcommand{\nbr} {\ensuremath{\langle ij \rangle}}
\newcommand{\ncap} {\ensuremath{\hat{n}}}

\begin{abstract}
  We propose a new method of optical lattice modulation spectroscopy
  for studying the spectral function of ultracold bosons in an optical
  lattice. We show that different features of the single particle
  spectral function in different quantum phases can be obtained by
  measuring the change in momentum distribution after the modulation. In the Mott phase, this gives information about
  the momentun dependent gap to particle-hole excitations as well as
  their spectral weight. In the
  superfluid phase, one can obtain the spectrum of the gapless
  Bogoliubov quasiparticles as well as the gapped amplitude
  fluctuations. The distinct evolution of the response with modulation
  frequency in the two phases can be used to identify these phases and the quantum phase transition separating them.
\end{abstract}

\title{Momentum Resolved Optical Lattice Modulation Spectroscopy for Bose Hubbard Model}
\author{Rajdeep Sensarma$^1$, K. Sengupta$^2$ and S. Das Sarma$^1$}
\affiliation{$1$ Condensed Matter Theory Center, University of
  Maryland, College Park, Maryland 20742, USA\\ $2$ Theoretical
Physics Department, Indian Association for the Cultivation of
Science, Kolkata 700032, India.}

\date{\today}

\maketitle

Ultracold atoms on optical lattices have opened up new possibilities
of studying interacting quantum systems with tunable parameters in a
controlled environment~\cite{blochreview}. A seminal achievement
in this field has been the implementation of the Bose Hubbard
model~\cite{jaksch98}, which is used as the paradigm for
superfluid-insulator (SI) transitions, and has been studied extensively
both analytically ~\cite{fisher89,sheshadri93,Sengupta05,hrk10}, and
numerically~\cite{krauth95,prokofiev03}. The observation of the SI
transition~\cite{greiner02} in this system is a successful milestone
in the study of ultracold atomic systems on optical lattices. More
complex but related systems like Bose-Fermi mixtures~\cite{BoseFermi},
spinor Bose condensates~\cite{SpinorBEC}, and Bose Hubbard model with
an artificial vector potential created by Raman beams~\cite{GaugeExpt}
have also been implemented leading to a host of novel phenomena
~\cite{Jaksch:Hofstadter,Powell,Sengupta:Gauge1,Sengupta:Gauge2}.

The SI transition in these systems is usually observed through time of
flight images of boson density which yields information about the
boson momentum distribution, $n_{\bf k}$, in the equilibrium state
~\cite{greiner02}. The superfluid phase is then identified by sharp
peaks of $n_\kk$ at zero momentum (and at corresponding Bragg
vectors), while the Mott phase shows a diffuse density pattern. The
contrast of the peaks (or visibility) has traditionally been used to
track such SI transitions.  This is, however, unsatisfactory for a
clean distinction between the two quantum phases. The main problem
with this method, as pointed out in earlier works \cite{Jason1}, is
that $n_{\bf k}$ has a precursor peak at ${\bf k}=0$ near the SI
transition ~\cite{Sengupta05}. Thus looking for a signature of
superfluidity involves looking for an extra peaky feature on top of
the generalized cluster around ${\bf k}=0$ which is experimentally
challenging. The other approach that has been suggested is to measure
the local density in situ with a very fine resolution and look for a
plateau of constant density corresponding to the Mott phase
~\cite{Jason2,Chin}. An alternate direct way of distinguishing between
the superfluid and Mott phases is to measure the excitation spectra in
these two phases. The superfluid phase has gapless spectrum with a
linear low energy dispersion, while the spectrum in the Mott phase is
gapped. Indeed the energy transfer due to optical lattice modulation
\cite{Esslinger,Huber,Sensarma} has been previously studied as a
possible probe for the Bose Hubbard model~\cite{Kollath,Ehud}.
However, such a method does not yield momentum-resolved information
about the spectral function of the bosons.

In this letter, we propose a novel method based on optical lattice
modulation~\cite{Esslinger,Huber,Sensarma} to obtain the single
particle spectral properties of the Bose-Hubbard model in both the
Mott and the superfluid phases. (In a broad qualitative sense, our
proposed technique, as we establish in this work, is somewhat akin to
being the bosonic counterpart of the ARPES technique used extensively
in solid state electronic materials in the sense that our method gives
direct momentum- and frequency-resolved spectral information about the
low-lying quantum excitations of the interacting system.)  Our
proposed experiment is the following: we turn on an additional weak
modulating optical lattice at frequency $\omega$ for some time before
turning off all lattice and trap potentials simultaneously and
measuring the density of the particles after a time of flight. For a
sufficiently long time of flight, the density in real space can be
mapped back to the momentum distribution of the system after the
lattice modulation. We propose to compare this momentum distribution
to that in the unperturbed case (without the modulation). We note at
the outset that such a comparison of momentum distributions involves
looking at the excitations created by the optical lattice modulation
and not measuring equilibrium $n_{\bf k}$ of the bosons. Hence the
proposed experiment does not suffer from the problem of measuring the
equilibrium $n_{\bf k}$ mentioned earlier. We show that the change in
the momentum distribution as a function of modulating frequency
$\omega$, $\delta n_{\bf k}(\omega)$, carries information about the
interacting spectral function both in the Mott and the superfluid side
of the phase diagram.  Using a non-perturbative strong coupling
expansion\cite{Sengupta05}, we compute $\delta n_{\bf k}(\omega)$ and
provide a precise algorithm to extract spectral information from the measurement. We demonstrate that in the Mott phase, the
variation of $\delta n_{\bf k}(\omega)$ in the Brillouin zone with
$\omega$ tracks the dispersion of the gapped particle-hole excitations
while in the superfluid phase, $\delta n_{\bf k}$ shows a gapless
linear dispersion at low $\omega$ and a gapped amplitude mode at high
$\omega$. This leads to a {\it qualitative} distinction between the
two phases. The height of the peaks of $\delta n_{\bf k}$ in the
Brillouin zone also contain important information about the spectral
weights of the different modes. Note that our proposed method contains detailed momentum resolved information which is not accessible by
looking at the total energy transferred by lattice modulation
suggested earlier \cite{Kollath,Ehud}. Recently the study of these
spectral features through Bragg spectroscopy (which measures the
structure factor) has been suggested and the corresponding response
functions have been calculated~\cite{Ehud}. However, compared to Bragg
spectroscopy, our method is simpler to implement as it does not
require additional lasers and the data can be collected in a single
shot. Additionally, our method can yield momentum resolved spectral
information, while Bragg spectroscopy integrates over the single
particle momenta. Thus our proposal constitutes a significant
improvement over the existing experimental proposals for both accurate
detection of the SI transition and for obtaining the momentum-resolved
spectral information for bosonic excitations in ultracold atom systems.

We consider an ultracold atomic implementation of a repulsive Bose
Hubbard model on a d-dimensional hypercubic optical lattice described by
the Hamiltonian
\beq
H=-J\sum_{\nbr} \ba_ib_j +U\sum_i n_i(n_i-1) -\mu\sum_i n_i,
\label{ham}
\eeq
where $\ba_i$ creates a boson on the site $i$ and $n_i=\ba_ib_i$ is
the number operator on site $i$, and we shall focus on $d=2,3$ in
the present work. Here $J$, $U$ and $\mu$ are respecively the
tunneling parameter, the onsite repulsion and the chemical potential
of the system. At $T=0$ this system undergoes a SI transition as a
function of $J/U$ (or density or equivalently $\mu/U$) with the Mott
insulating phase occuring at small $J/U$ and at a commensurate
integer density on each site. On top of this system, we consider a
small perturbation to the optical lattice potential modulated at a
frequency $\omega$. Since the tunneling is exponentially dependent
on the height of the optical lattice while other parameters have
much weaker dependence, the primary effect of this perturbation is
to modulate the tunneling parameter. The perturbing
Hamiltonian can then be written as
\beq
H_1=-\lambda J\cos(\omega t)\sum_{\nbr} \ba_ib_j,
\label{pert}
\eeq
where $t$ denotes time and $\lambda$ is a small perturbation
parameter. For a deep lattice, $J \sim E_Rexp(-\sqrt{V_0/E_R})$ where
$V_0$ is the lattice height and $E_R$ is the recoil energy. Hence
$\lambda\sim \sqrt{V_0/E_R}\delta V_0/V_0$, where $\delta V_0$ is the
amplitude of modulation of the optical lattice height. This gives an order
of magnitude for $\lambda$. More quantitatively accurate estimates can be
easily obtained by solving the band structure problem.

%Our proposed experiment is the following: After the optical lattice
%is modulated for some time, both the perturbation and the lattice
%potential is switched off simultaneously and after a time of flight
%the density of the particles is measured. For a sufficiently long
%time of flight, the density in real space can be mapped back to the
%momentum distribution of the system after the lattice modulation. In
%contrast to the standard time of flight experiments, here we propose
%to compare this momentum distribution to that in the unperturbed
%case. In the rest of this work, we show that the change in the
%momentum distribution as a function of modulating frequency, $\delta
%n_{\bf k}(\omega)$, carries information about the spectrum of the
%model both in the Mott insulator and the superfluid side of the
%phase diagram. We develop a theory for computing $\delta n_{\bf
%k}(\omega)$ and provide an algorithm to extract the spectral
%information of the model from it in an measurement.

\begin{figure}[t]
\centering
\includegraphics[width=0.5\textwidth]{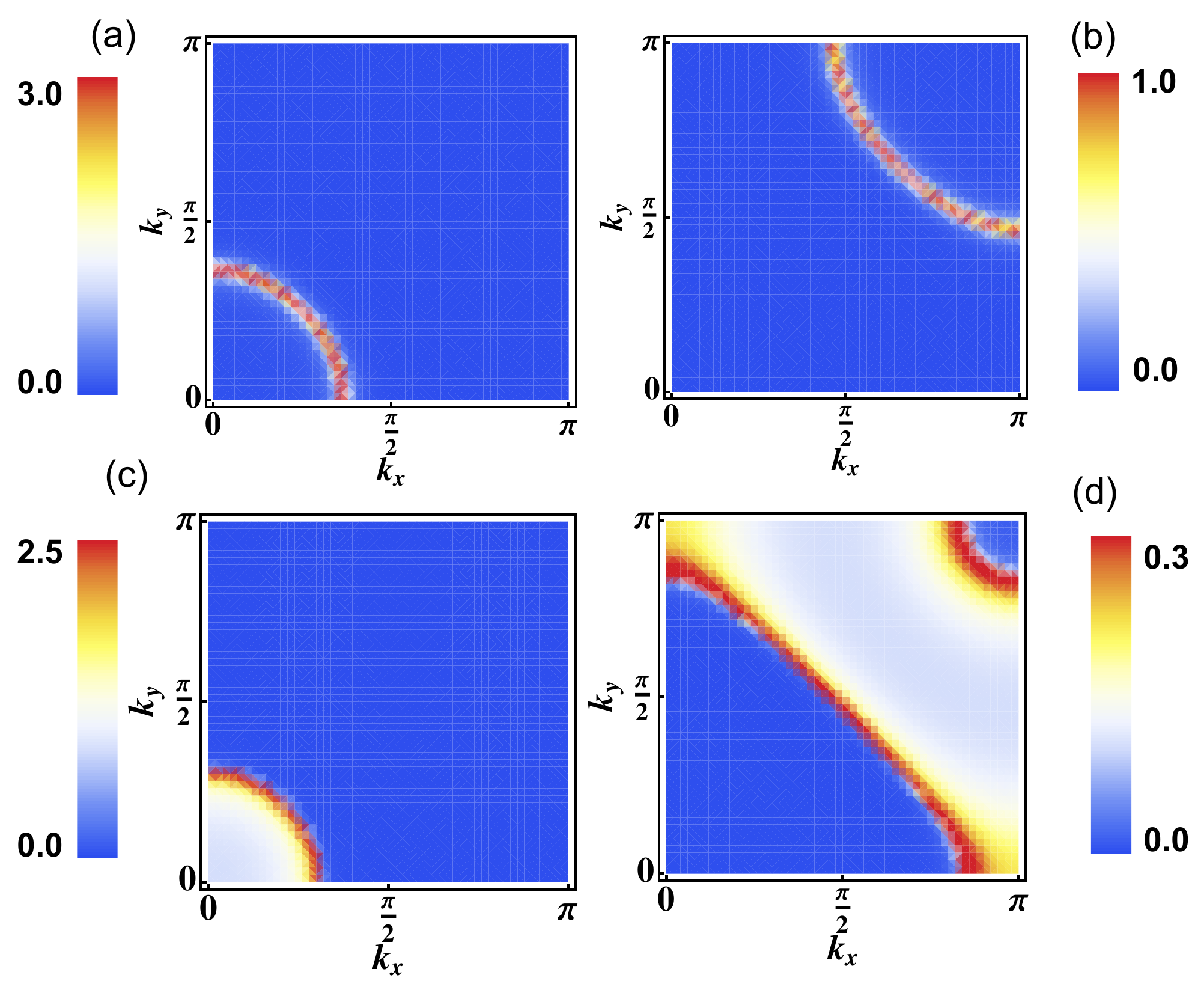}
\caption{ The out of phase response of the momentum distribution in
the $n=1$ Mott phase of 2D Bose Hubbard model ($U=30J$ and
$\mu=20J$) for (a)$\omega=20 J$
and (b) $\omega=35J$. (c) and (d): Integrated response from 3D Bose Hubbard
model ($U=40J$ and $\mu=10.5J$) for (c) $\omega=20 J$ and (b) $\omega=45J$. The perturbation parameter
$\lambda=0.1$ for all the plots. (Color online)}
\label{mottresponse:bz}
\end{figure}
%The presence of a condensate (i.e. of anomalous propagators in Green
%function language) in the superfluid phase makes the theory slightly
%more involved from that in the Mott phase. So, we will first focus

We begin with the Mott phase of the model. Within
linear response (linear in $\lambda$), the change in $n_\kk$ is
\beq
\delta n_\kk(t)=n_1(\kk,\omega) \cos \omega t +n_2(\kk,\omega)
\sin \omega t,
\label{deltank}
\eeq
where the \emph {in phase} and \emph{out
of phase} response of the system, $n_1$ and $n_2$, are given by the
real and imaginary part of the response function
$\Pi(\kk,\omega+i0^+)$ with
\beq
\label{pik}
\Pi(\kk,i\omega)=-\frac{1}{\beta}\ep_\kk \sum_{i\omega_n} G(\kk,i\omega_n)G(\kk,i\omega+i\omega_n).
\eeq
Here $\beta=1/T$, $T$ being the temperature of the system,
$\ep_\kk=-2J \sum_{i=1,d}\cos k_i$ is the bare band dispersion of
the system and $G(\kk,\omega)$ is the single-particle Green function
of the system. The response function in eq.~\ref{pik} neglects
vertex corrections although self-energy corrections can be included
in the Green functions used to evaluate it.  Note that for a
non-interacting system $n_2$ is identically zero as the perturbation commutes with the non-interacting
Hamiltonian.

The single particle
Green function in the Mott phase within the random phase approximation
(RPA)~\cite{Sengupta05} is given by
\beq
%G(\kk,i\omega)=\frac{z_\kk}{i\omega-E^+_\kk}+\frac{1-z_\kk}{i\omega-E^-_\kk}
G(\kk,i\omega)=z_\kk(i\omega-E^+_\kk)^{-1}
+(1-z_\kk)(i\omega-E^-_\kk)^{-1} ,
\label{gkw} \eeq
where the particle (hole) dispersions are given by
$E^{+(-)}_\kk=-\delta \mu+\frac{1}{2}\left(\ep_\kk\pm\sqrt{\ep_\kk^2+4\ep_\kk Ux+U^2}\right)$
and the spectral residue is given by
$z_\kk=(E^+_\kk+\delta \mu+Ux)/(E^+_\kk-E^-_\kk)$.
Here $x=n+1/2$ and $\delta \mu=\mu-U(n-1/2)$ where $n$ is the integer
number of bosons on each site. Using the above form of the boson Green function, we then obtain
\bqa
\no n_1(\kk,\omega)& = &2\lambda\ep_\kk \frac{\Delta_\kk z_\kk(1-z_\kk)[n_B(E^-_\kk)-n_B(E^+_\kk)]}{\omega^2-\Delta_\kk^2},\\
\no n_2(\kk,\omega)&=&\pi \lambda\ep_\kk z_\kk(1-z_\kk)[n_B(E^-_\kk)-n_B(E^+_\kk)]\\
& &\times [\delta(\omega+\Delta_\kk)-\delta(\omega-\Delta_\kk)],
\label{nkw}
\eqa
where $\Delta_\kk=E^+_\kk-E^-_\kk$ is the momentum dependent gap to
creating particle-hole excitations in the system with zero center of
mass momentum and $n_B$ is the Bose distribution function. It is
clear from the form of $\Delta_\kk$ that it has the lowest (but
finite) value $\Delta_M$ at the zone center ($\kk=[0,0]$) and
increases as we go out towards the edge of the Brillouin zone. The SI transition is then marked by the vanishing of $\Delta_M$

We focus here on $n_2(\kk,\omega)$, which shows sharp features in
the response. For modulation frequencies below the Mott gap,
$n_2(\kk,\omega)$ is zero everywhere in the Brillouin zone. For
frequencies above the Mott gap $n_2$ will have response at the
contours corresponding to $\Delta_\kk=\omega$ with a weight
proportional to $z_\kk(1-z_\kk)$. Looking at $n_1(\kk,\omega)$, we
see that for frequencies less than the Mott gap, the response is
broad with a broad peak at the zone center. For frequencies above
the Mott gap, the response is sharply peaked around the contours of
$\Delta_\kk=\omega$ and changes sign as this contour is crossed.
This qualitative reasoning is verified in
Fig.~\ref{mottresponse:bz}(a) and (b), where $n_2(\kk,\omega)$ is
plotted as a function of $k_x$ and $k_y$ for $d=2$ and two different
frequencies corresponding to (a) $\Delta_M< \omega<U$ and (b)
$\omega >U$. For frequencies above the Mott gap, we see a contour of
excitations which moves from the zone center to the edge of the
Brillouin zone as the frequency is increased. Thus, one can
determine the momentum dependent gap by doing a frequency sweep and
looking at the peaks in the response. We note that for the 3D
Bose-Hubbard model, the measured momentum distribution is
integrated along one axis and hence an analogous plot will reflect
$\int_{-\pi}^{\pi} dk_z n_2({\bf k}, \omega)/2 \pi$. The column
integration leads to a region of excitations with the perimeter
defined by the contour of excitations in the $k_x-k_y$ plane. This
is plotted in Fig.~\ref{mottresponse:bz} (c) and (d), for two
different values of $\omega$.
\begin{figure}[t!]
\centering
\includegraphics[width=0.5\textwidth]{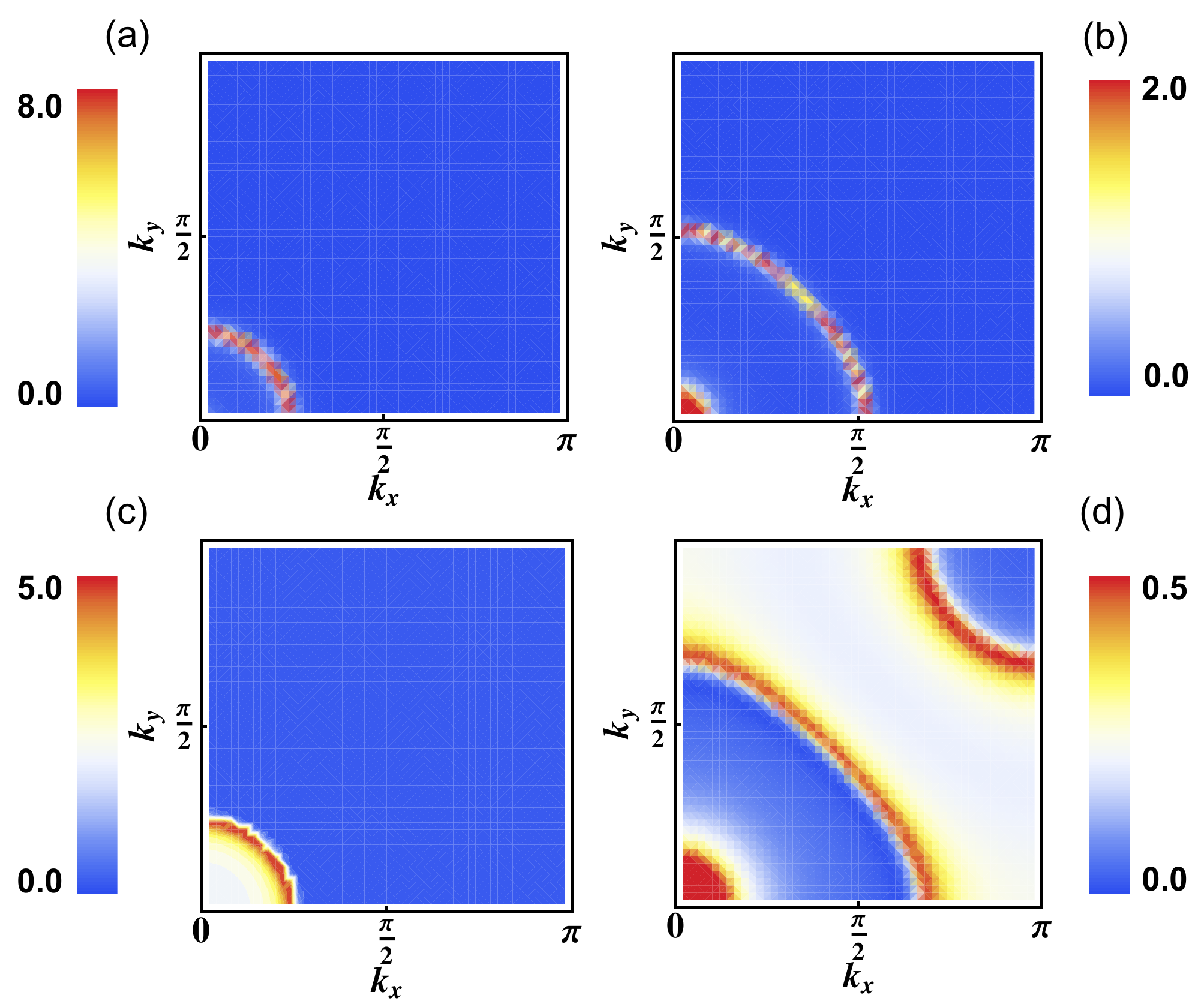}
\caption{(a) and (b) : The out of phase response of the momentum distribution in the
superfluid phase for a 2D Bose Hubbard model ($U=5J$ and
$\mu=2.3J$) for (a) $\omega=4J$ (b) $\omega=7J$. In (a) the response is due to creation of two phase
modes. In (b) the low $\kk$ response is from simultaneous excitation
of a phase and an amplitude mode. There is also a small response
from excitation of two phase modes at large $\kk$. (c) and (d): Column integrated out of phase response for 3D Bose Hubbard model ($U=8J$ and
$\mu=2.2J$) for  (c) $\omega=4.2J$ (b) $\omega=11.2J$.  The rings seen in the 2D response are now filled up due to column integration. The perturbation parameter $\lambda=0.1$ for all plots.(Color online) } \label{fig:sfresponse}
\end{figure}

We now turn our attention to the superfluid phase, which is
characterized by the presence of a macroscopic condensate and finite
expectation values of the anomalous propagators $F(\kk,i\omega_n)=
\langle b^\dagger_\kk b^\dagger_{-\kk}\rangle$. In this case, the response
function $\Pi(\kk,i\omega)$ in Eq.~\ref{pik} has an extra term
coming from the anomalous propagators and is given by
\bqa
\no \Pi(\kk,i\omega_n)& = &-\frac{1}{\beta}\ep_\kk \sum_{i\omega_n} \big[G(\kk,i\omega_n)G(\kk,i\omega+i\omega_n)\\
& + & F(\kk,i\omega_n)F(\kk,i\omega-i\omega_n) \big]. \eqa
In the superfluid phase, the normal and the anomalous Green
functions are given by~\cite{Sengupta05}
\bqa
\no G(\kk,i\omega_n)&=& \frac{g_{1\kk}}{i\omega_n-E_{1\kk}}+\frac{g_{2\kk}}{i\omega_n+E_{1\kk}}\\
&&+\frac{g_{3\kk}}{i\omega_n-E_{2\kk}}+\frac{g_{4\kk}}{i\omega_n+E_{2\kk}},
\nonumber\\
F(\kk,i\omega_n)&=&
\Phi\left[\frac{f_{\kk}}{i\omega_n^2-E_{1\kk}^2}+\frac{1-f_{\kk}}{i\omega_n^2-E_{2\kk}^2}\right],
\eqa
where $E^2_{1(2)\kk}=-B_\kk\pm\sqrt{B_\kk^2-C_\kk}$ with
$B_\kk=b_\kk-a_\kk^2/2+\Phi^2/2$ and
$C_\kk=b_\kk^2-\Phi^2(\mu+U)^2$. Here $a_\kk =
2\delta\mu-\ep_\kk-2\Phi$, $\Phi =
\frac{\delta\mu^2-U^2/4}{\mu+U}+zJ$, and $b_\kk = \delta\mu^2
-U^2/4-(\ep_\kk+2\Phi)(\mu+U)$, with $z=2d$ for a hypercubic lattice in d
dimensions. The quasiparticle weights are given by
$g_{1(2)\kk}=[E_{1\kk}\pm (\mu+U)][E_{1\kk}^2\mp a_\kk
E_{1\kk}+b_\kk]/(2E_{1\kk}[E_{1\kk^2}^2-E_{2\kk}^2])$, $g_{3(4)\kk}$
is obtained from $g_{1(2)\kk}$ by taking $E_{1\kk}\leftrightarrow
E_{2\kk}$, and
$f_{\kk}=(E_{1\kk}^2-(\mu+U)^2)/(E_{1\kk}^2-E_{2\kk}^2)$. The pole
$E_{2\kk}$ is the gapless phase fluctuation mode at small
$\kk$ (near the zone center). The pole $E_{1\kk}$ represents the
amplitude mode at low $\kk$ and is gapped. At large $\kk$ the
amplitude and phase modes are mixed; however, we still refer to these
modes as amplitude and phase mode for easy identification. The phase
mode carries almost all the spectral weight at low $\kk$, while the
spectral weight resides mainly in the gapped mode at larger $\kk$
values.

Using these forms for the propagators we can now evaluate the
momentum distribution response function. For simplicity, we restrict
ourselves to $T=0$ and $\omega>0$ (note that the in-phase and out of
phase responses are respectively symmetric and antisymmetric in
$\omega$). Then, the change in momentum distribution is given by
\bqa
\no n_1(\kk,\omega)& = &2\pi\lambda \ep_\kk \left[\nu_\kk\frac{E_{1\kk}+E_{2\kk}}{\omega^2-(E_{1\kk}+E_{2\kk})^2}\right. \\
\no & & \left. +\alpha_\kk \frac{2E_{1\kk}}{\omega^2-4E_{1\kk}^2}+\beta_\kk \frac{2E_{2\kk}}{\omega^2-4E_{2\kk}^2}\right],\\
n_2(\kk,\omega)& = &\pi\lambda \ep_\kk \left[\nu_\kk\delta(\omega-E_{1\kk}-E_{2\kk})\right. \\
\no & & \left. +\alpha_\kk \delta(\omega-2E_{1\kk})+\beta_\kk
\delta(\omega-2E_{2\kk})\right], \label{reseq1} \eqa
where
$\nu_\kk=-g_{1\kk}g_{4\kk}-g_{2\kk}g_{3\kk}+\Phi^2f_\kk(1-f_\kk)/(2E_{1\kk}E_{2\kk})$,
$\alpha_\kk=-g_{1\kk}g_{2\kk}+\Phi^2f_\kk^2/(4E_{1\kk}^2)$, and
$\beta_\kk=-g_{3\kk}g_{4\kk}+\Phi^2(1-f_\kk)^2/(4E_{2\kk}^2)$.

Eq.\ \ref{reseq1} shows that the out of phase response in the SF
phase is peaked at energies corresponding to the creation of (a) two
phase modes (b) an amplitude and a phase mode and (c) two amplitude modes. The in-phase response also shows broader peaks at these
contours. At low frequencies (less than the gap in $E_{1\kk}$), the
experiments should see a single contour of excitations corresponding
to two phase mode excitations, as shown in
Fig~\ref{fig:sfresponse}(a) for $d=2$. As the frequency crosses the
gap of the amplitude mode, two contours should be visible, a large
contour corresponding to two phase modes and a small $\kk$ contour
corresponding to an amplitude and a phase mode, as in
Fig~\ref{fig:sfresponse}(b).  As the frequency crosses twice the
gap, the two phase modes disappear as the frequency is above their
bandwidth. The large $\kk$ response is due to one amplitude and one
phase mode. There is a low $\kk$ response from two amplitude modes,
but it is heavily suppressed due to spectral weights. For $d=3$, a
similar pattern is observed. However, the contours are replaced by
regions with boundaries given by the location of the corresponding
excitations in the $k_x-k_y$ plane, as shown in
Fig.~\ref{fig:sfresponse}(c) and (d).

Thus we see that the change in momentum distribution due to optical
lattice modulation can be used to obtain the excitation spectrum in
both the Mott and the superfluid phase. In the Mott phase, the
particle-hole spectrum is gapped and there is no low frequency
response. For both $d=2$ and $3$ and for modulation frequencies
above the Mott gap, the response is peaked at the contour in the
Brillouin zone whose dispersion matches the modulation frequency.
Thus the peak response occurs on contours of increasing size as the
frequency is increased. On the other hand, in the superfluid phase,
the existence of gapless phase modes leads to low frequency response
on contours whose size increase with frequency. However, on crossing
the gap for the amplitude mode, there is a large response near the
zone center and two distinct contours can be seen in this regime. At
higher frequencies, there is a third contour of excitations
corresponding to two amplitude modes, where the response is suppressed due to low spectral weight.

In conclusion, we have derived the response of the momentum
distribution of a Bose Hubbard model to optical lattice modulation
within a strong coupling expansion,
both in the Mott insulator and the superfluid phase. In real
experiments, the sharp poles in the single particle
Green functions will be smeared due to several factors leading to a
broadening of the response shown. The harmonic trap
will smear the distribution on a momentum scale of the order of the
inverse oscillator length scale. For a wide trap, this is usually
negligible compared to other sources of smearing. Finite temperature
will lead to a smearing on energy scales $\sim T$. For the Mott
insulating phase, this smearing would be negligible as long as the
temperature is much smaller than the Mott gap ($T\ll U$)
\cite{Gerbier07,Weld09}, which is easily achievable experimentally. For the
superfluid phase, the low energy excitations would be smeared and it
would be impossible to follow the very long wavelength low energy
response (below energy $\sim T$). However, the strong momentum
dependence of the spectral weight of the phase modes would still
lead to relatively well defined peaks in the Brillouin zone. On the
other hand, deep in the superfluid phase, the amplitude modes are
gapped and the emergence of the small contour around the gap
frequency should be easy to observe. Finally, including vertex
corrections on top of the random phase approximation would broaden
the poles, especially for the amplitude modes. The broadening
represents processes where the amplitude mode can decay into two or
more phase modes. However, for lattice systems, the phase modes have
a finite bandwidth and a large gap for the amplitude mode restricts
the phase space for such processes. Further, the two phonon decay
involves phonon modes at relatively large energies which carry very
little spectral weight. Thus the amplitude modes would be sharply
defined near the zone center and hence the emergence of the small
response contour around the gap should be easily observable.

R.S. and S.D.S. acknowledges support from ARO-DARPA-OLE, ARO-MURI and
JQI-NSF-PFC.  K.S. thanks DST, India for support through project
no. SR/S2/CMP-001/2009.

\bibliography{bh_mod}

\begin{thebibliography}{26}
\expandafter\ifx\csname natexlab\endcsname\relax\def\natexlab#1{#1}\fi
\expandafter\ifx\csname bibnamefont\endcsname\relax
  \def\bibnamefont#1{#1}\fi
\expandafter\ifx\csname bibfnamefont\endcsname\relax
  \def\bibfnamefont#1{#1}\fi
\expandafter\ifx\csname citenamefont\endcsname\relax
  \def\citenamefont#1{#1}\fi
\expandafter\ifx\csname url\endcsname\relax
  \def\url#1{\texttt{#1}}\fi
\expandafter\ifx\csname urlprefix\endcsname\relax\def\urlprefix{URL }\fi
\providecommand{\bibinfo}[2]{#2}
\providecommand{\eprint}[2][]{\url{#2}}

\bibitem[{\citenamefont{Bloch et~al.}(2008)\citenamefont{Bloch, Dalibard, and
  Zwerger}}]{blochreview}
\bibinfo{author}{\bibfnamefont{I.}~\bibnamefont{Bloch}},
  \bibinfo{author}{\bibfnamefont{J.}~\bibnamefont{Dalibard}}, \bibnamefont{and}
  \bibinfo{author}{\bibfnamefont{W.}~\bibnamefont{Zwerger}},
  \bibinfo{journal}{Rev. Mod. Phys.} \textbf{\bibinfo{volume}{80}},
  \bibinfo{pages}{885} (\bibinfo{year}{2008}).

\bibitem[{\citenamefont{Jaksch et~al.}(1998)\citenamefont{Jaksch, Bruder,
  Cirac, Gardiner, and Zoller}}]{jaksch98}
\bibinfo{author}{\bibfnamefont{D.}~\bibnamefont{Jaksch}},
  \bibinfo{author}{\bibfnamefont{C.}~\bibnamefont{Bruder}},
  \bibinfo{author}{\bibfnamefont{J.~I.} \bibnamefont{Cirac}},
  \bibinfo{author}{\bibfnamefont{C.~W.} \bibnamefont{Gardiner}},
  \bibnamefont{and} \bibinfo{author}{\bibfnamefont{P.}~\bibnamefont{Zoller}},
  \bibinfo{journal}{Phys. Rev. Lett.} \textbf{\bibinfo{volume}{81}},
  \bibinfo{pages}{3108} (\bibinfo{year}{1998}).

\bibitem[{\citenamefont{Fisher et~al.}(1989)\citenamefont{Fisher, Weichman,
  Grinstein, and Fisher}}]{fisher89}
\bibinfo{author}{\bibfnamefont{M.~P.~A.} \bibnamefont{Fisher}},
  \bibinfo{author}{\bibfnamefont{P.~B.} \bibnamefont{Weichman}},
  \bibinfo{author}{\bibfnamefont{G.}~\bibnamefont{Grinstein}},
  \bibnamefont{and} \bibinfo{author}{\bibfnamefont{D.~S.}
  \bibnamefont{Fisher}}, \bibinfo{journal}{Phys. Rev. B}
  \textbf{\bibinfo{volume}{40}}, \bibinfo{pages}{546} (\bibinfo{year}{1989}).

\bibitem[{\citenamefont{Sheshadri et~al.}(1993)\citenamefont{Sheshadri,
  Krishnamurthy, Pandit, and Ramakrishnan}}]{sheshadri93}
\bibinfo{author}{\bibfnamefont{K.}~\bibnamefont{Sheshadri}},
  \bibinfo{author}{\bibfnamefont{H.}~\bibnamefont{Krishnamurthy}},
  \bibinfo{author}{\bibfnamefont{R.}~\bibnamefont{Pandit}}, \bibnamefont{and}
  \bibinfo{author}{\bibfnamefont{T.}~\bibnamefont{Ramakrishnan}},
  \bibinfo{journal}{Euro. Phys. Lett.} \textbf{\bibinfo{volume}{22}},
  \bibinfo{pages}{257} (\bibinfo{year}{1993}).

\bibitem[{\citenamefont{Sengupta and Dupuis}(2005)}]{Sengupta05}
\bibinfo{author}{\bibfnamefont{K.}~\bibnamefont{Sengupta}} \bibnamefont{and}
  \bibinfo{author}{\bibfnamefont{N.}~\bibnamefont{Dupuis}},
  \bibinfo{journal}{Phys. Rev. A} \textbf{\bibinfo{volume}{71}},
  \bibinfo{pages}{033629} (\bibinfo{year}{2005}).

\bibitem[{\citenamefont{Freericks et~al.}(2009)\citenamefont{Freericks,
  Krishnamurthy, Kato, Kawashima, and Trivedi}}]{hrk10}
\bibinfo{author}{\bibfnamefont{J.~K.} \bibnamefont{Freericks}},
  \bibinfo{author}{\bibfnamefont{H.~R.} \bibnamefont{Krishnamurthy}},
  \bibinfo{author}{\bibfnamefont{Y.}~\bibnamefont{Kato}},
  \bibinfo{author}{\bibfnamefont{N.}~\bibnamefont{Kawashima}},
  \bibnamefont{and} \bibinfo{author}{\bibfnamefont{N.}~\bibnamefont{Trivedi}},
  \bibinfo{journal}{Phys. Rev. A} \textbf{\bibinfo{volume}{79}},
  \bibinfo{pages}{053631} (\bibinfo{year}{2009}).

\bibitem[{\citenamefont{Krauth and Trivedi}(1991)}]{krauth95}
\bibinfo{author}{\bibfnamefont{W.}~\bibnamefont{Krauth}} \bibnamefont{and}
  \bibinfo{author}{\bibfnamefont{N.}~\bibnamefont{Trivedi}},
  \bibinfo{journal}{Europhys. Lett.} \textbf{\bibinfo{volume}{14}},
  \bibinfo{pages}{627} (\bibinfo{year}{1991}).

\bibitem[{\citenamefont{Capogrosso-Sansone
  et~al.}(2007)\citenamefont{Capogrosso-Sansone, Prokof'ev, and
  Svistunov}}]{prokofiev03}
\bibinfo{author}{\bibfnamefont{B.}~\bibnamefont{Capogrosso-Sansone}},
  \bibinfo{author}{\bibfnamefont{N.~V.} \bibnamefont{Prokof'ev}},
  \bibnamefont{and} \bibinfo{author}{\bibfnamefont{B.~V.}
  \bibnamefont{Svistunov}}, \bibinfo{journal}{Phys. Rev. B}
  \textbf{\bibinfo{volume}{75}}, \bibinfo{pages}{134302}
  (\bibinfo{year}{2007}).

\bibitem[{\citenamefont{Greiner et~al.}(2002)\citenamefont{Greiner, Mandel,
  Esslinger, Hansch, and Bloch}}]{greiner02}
\bibinfo{author}{\bibfnamefont{M.}~\bibnamefont{Greiner}},
  \bibinfo{author}{\bibfnamefont{O.}~\bibnamefont{Mandel}},
  \bibinfo{author}{\bibfnamefont{T.}~\bibnamefont{Esslinger}},
  \bibinfo{author}{\bibfnamefont{T.~W.} \bibnamefont{Hansch}},
  \bibnamefont{and} \bibinfo{author}{\bibfnamefont{I.}~\bibnamefont{Bloch}},
  \bibinfo{journal}{Nature} \textbf{\bibinfo{volume}{415}}, \bibinfo{pages}{39}
  (\bibinfo{year}{2002}).

\bibitem[{\citenamefont{G\"unter et~al.}(2006)\citenamefont{G\"unter,
  St\"oferle, Moritz, K\"ohl, and Esslinger}}]{BoseFermi}
\bibinfo{author}{\bibfnamefont{K.}~\bibnamefont{G\"unter}},
  \bibinfo{author}{\bibfnamefont{T.}~\bibnamefont{St\"oferle}},
  \bibinfo{author}{\bibfnamefont{H.}~\bibnamefont{Moritz}},
  \bibinfo{author}{\bibfnamefont{M.}~\bibnamefont{K\"ohl}}, \bibnamefont{and}
  \bibinfo{author}{\bibfnamefont{T.}~\bibnamefont{Esslinger}},
  \bibinfo{journal}{Phys. Rev. Lett.} \textbf{\bibinfo{volume}{96}},
  \bibinfo{pages}{180402} (\bibinfo{year}{2006}).

\bibitem[{\citenamefont{Sadler et~al.}(2006)\citenamefont{Sadler, Higbie,
  Leslie, Vengalattore, and Stamper-Kurn}}]{SpinorBEC}
\bibinfo{author}{\bibfnamefont{L.~E.} \bibnamefont{Sadler}},
  \bibinfo{author}{\bibfnamefont{J.~M.} \bibnamefont{Higbie}},
  \bibinfo{author}{\bibfnamefont{S.~R.} \bibnamefont{Leslie}},
  \bibinfo{author}{\bibfnamefont{M.}~\bibnamefont{Vengalattore}},
  \bibnamefont{and}
  \bibinfo{author}{\bibfnamefont{D.}~\bibnamefont{Stamper-Kurn}},
  \bibinfo{journal}{Nature} \textbf{\bibinfo{volume}{443}},
  \bibinfo{pages}{312} (\bibinfo{year}{2006}).

\bibitem[{\citenamefont{Lin et~al.}(2009)\citenamefont{Lin, Compton,
  Jimenez-Garcia, Porto, and Spielman}}]{GaugeExpt}
\bibinfo{author}{\bibfnamefont{Y.-J.} \bibnamefont{Lin}},
  \bibinfo{author}{\bibfnamefont{R.~L.} \bibnamefont{Compton}},
  \bibinfo{author}{\bibfnamefont{K.}~\bibnamefont{Jimenez-Garcia}},
  \bibinfo{author}{\bibfnamefont{J.~V.} \bibnamefont{Porto}}, \bibnamefont{and}
  \bibinfo{author}{\bibfnamefont{I.~B.} \bibnamefont{Spielman}},
  \bibinfo{journal}{Nature} \textbf{\bibinfo{volume}{462}},
  \bibinfo{pages}{628} (\bibinfo{year}{2009}).

\bibitem[{\citenamefont{Jaksch}(2008)}]{Jaksch:Hofstadter}
\bibinfo{author}{\bibfnamefont{D.}~\bibnamefont{Jaksch}},
  \bibinfo{journal}{New. Journ. Phys.} \textbf{\bibinfo{volume}{111}},
  \bibinfo{pages}{000} (\bibinfo{year}{2008}).

\bibitem[{\citenamefont{Powell et~al.}(2010)\citenamefont{Powell, Barnett,
  Sensarma, and Das~Sarma}}]{Powell}
\bibinfo{author}{\bibfnamefont{S.}~\bibnamefont{Powell}},
  \bibinfo{author}{\bibfnamefont{R.}~\bibnamefont{Barnett}},
  \bibinfo{author}{\bibfnamefont{R.}~\bibnamefont{Sensarma}}, \bibnamefont{and}
  \bibinfo{author}{\bibfnamefont{S.}~\bibnamefont{Das~Sarma}},
  \bibinfo{journal}{Phys. Rev. Lett.} \textbf{\bibinfo{volume}{104}},
  \bibinfo{pages}{255303} (\bibinfo{year}{2010}).

\bibitem[{\citenamefont{Sinha and Sengupta}(2010)}]{Sengupta:Gauge1}
\bibinfo{author}{\bibfnamefont{S.}~\bibnamefont{Sinha}} \bibnamefont{and}
  \bibinfo{author}{\bibfnamefont{K.}~\bibnamefont{Sengupta}},
  \bibinfo{journal}{arXiv:} p. \bibinfo{pages}{1003.0258}
  (\bibinfo{year}{2010}).

\bibitem[{\citenamefont{Saha et~al.}(2010)\citenamefont{Saha, Sengupta, and
  Ray}}]{Sengupta:Gauge2}
\bibinfo{author}{\bibfnamefont{K.}~\bibnamefont{Saha}},
  \bibinfo{author}{\bibfnamefont{K.}~\bibnamefont{Sengupta}}, \bibnamefont{and}
  \bibinfo{author}{\bibfnamefont{K.}~\bibnamefont{Ray}},
  \bibinfo{journal}{Phys. Rev. B} \textbf{\bibinfo{volume}{82}},
  \bibinfo{pages}{205126} (\bibinfo{year}{2010}).

\bibitem[{\citenamefont{Diener et~al.}(2007)\citenamefont{Diener, Zhou, Zhai,
  and Ho}}]{Jason1}
\bibinfo{author}{\bibfnamefont{R.~B.} \bibnamefont{Diener}},
  \bibinfo{author}{\bibfnamefont{Q.}~\bibnamefont{Zhou}},
  \bibinfo{author}{\bibfnamefont{H.}~\bibnamefont{Zhai}}, \bibnamefont{and}
  \bibinfo{author}{\bibfnamefont{T.-L.} \bibnamefont{Ho}},
  \bibinfo{journal}{Phys. Rev. Lett.} \textbf{\bibinfo{volume}{98}},
  \bibinfo{pages}{180404} (\bibinfo{year}{2007}).

\bibitem[{\citenamefont{Ho and Zhou}(2009)}]{Jason2}
\bibinfo{author}{\bibfnamefont{T.-L.} \bibnamefont{Ho}} \bibnamefont{and}
  \bibinfo{author}{\bibfnamefont{Q.}~\bibnamefont{Zhou}},
  \bibinfo{journal}{arXiv} \textbf{\bibinfo{volume}{0901.0018}}
  (\bibinfo{year}{2009}).

\bibitem[{\citenamefont{Gemelke et~al.}(2009)\citenamefont{Gemelke, Zhang,
  Hung, and Chin}}]{Chin}
\bibinfo{author}{\bibfnamefont{N.}~\bibnamefont{Gemelke}},
  \bibinfo{author}{\bibfnamefont{X.}~\bibnamefont{Zhang}},
  \bibinfo{author}{\bibfnamefont{C.-L.} \bibnamefont{Hung}}, \bibnamefont{and}
  \bibinfo{author}{\bibfnamefont{C.}~\bibnamefont{Chin}},
  \bibinfo{journal}{Nature} \textbf{\bibinfo{volume}{460}},
  \bibinfo{pages}{995} (\bibinfo{year}{2009}).

\bibitem[{\citenamefont{J\"ordens et~al.}(2008)\citenamefont{J\"ordens,
  Strohmaier, G\"unter, Moritz, and Esslinger}}]{Esslinger}
\bibinfo{author}{\bibfnamefont{R.}~\bibnamefont{J\"ordens}},
  \bibinfo{author}{\bibfnamefont{N.}~\bibnamefont{Strohmaier}},
  \bibinfo{author}{\bibfnamefont{K.}~\bibnamefont{G\"unter}},
  \bibinfo{author}{\bibfnamefont{H.}~\bibnamefont{Moritz}}, \bibnamefont{and}
  \bibinfo{author}{\bibfnamefont{T.}~\bibnamefont{Esslinger}},
  \bibinfo{journal}{Nature} \textbf{\bibinfo{volume}{455}},
  \bibinfo{pages}{204} (\bibinfo{year}{2008}).

\bibitem[{\citenamefont{Huber and R\"uegg}(2009)}]{Huber}
\bibinfo{author}{\bibfnamefont{S.~D.} \bibnamefont{Huber}} \bibnamefont{and}
  \bibinfo{author}{\bibfnamefont{A.}~\bibnamefont{R\"uegg}},
  \bibinfo{journal}{Phys. Rev. Lett.} \textbf{\bibinfo{volume}{102}},
  \bibinfo{pages}{065301} (\bibinfo{year}{2009}).

\bibitem[{\citenamefont{Sensarma et~al.}(2009)\citenamefont{Sensarma, Pekker,
  Lukin, and Demler}}]{Sensarma}
\bibinfo{author}{\bibfnamefont{R.}~\bibnamefont{Sensarma}},
  \bibinfo{author}{\bibfnamefont{D.}~\bibnamefont{Pekker}},
  \bibinfo{author}{\bibfnamefont{M.~D.} \bibnamefont{Lukin}}, \bibnamefont{and}
  \bibinfo{author}{\bibfnamefont{E.}~\bibnamefont{Demler}},
  \bibinfo{journal}{Phys. Rev. Lett.} \textbf{\bibinfo{volume}{103}},
  \bibinfo{pages}{035303} (\bibinfo{year}{2009}).

\bibitem[{\citenamefont{Kollath et~al.}(2006)\citenamefont{Kollath, Iucci,
  Giamarchi, Hofstetter, and Schollw\"ock}}]{Kollath}
\bibinfo{author}{\bibfnamefont{C.}~\bibnamefont{Kollath}},
  \bibinfo{author}{\bibfnamefont{A.}~\bibnamefont{Iucci}},
  \bibinfo{author}{\bibfnamefont{T.}~\bibnamefont{Giamarchi}},
  \bibinfo{author}{\bibfnamefont{W.}~\bibnamefont{Hofstetter}},
  \bibnamefont{and}
  \bibinfo{author}{\bibfnamefont{U.}~\bibnamefont{Schollw\"ock}},
  \bibinfo{journal}{Phys. Rev. Lett.} \textbf{\bibinfo{volume}{97}},
  \bibinfo{pages}{050402} (\bibinfo{year}{2006}).

\bibitem[{\citenamefont{Huber et~al.}(2007)\citenamefont{Huber, Altman,
  B\"uchler, and Blatter}}]{Ehud}
\bibinfo{author}{\bibfnamefont{S.~D.} \bibnamefont{Huber}},
  \bibinfo{author}{\bibfnamefont{E.}~\bibnamefont{Altman}},
  \bibinfo{author}{\bibfnamefont{H.~P.} \bibnamefont{B\"uchler}},
  \bibnamefont{and} \bibinfo{author}{\bibfnamefont{G.}~\bibnamefont{Blatter}},
  \bibinfo{journal}{Phys. Rev. B} \textbf{\bibinfo{volume}{75}},
  \bibinfo{pages}{085106} (\bibinfo{year}{2007}).

\bibitem[{\citenamefont{Gerbier}(2007)}]{Gerbier07}
\bibinfo{author}{\bibfnamefont{F.}~\bibnamefont{Gerbier}},
  \bibinfo{journal}{Phys. Rev. Lett.} \textbf{\bibinfo{volume}{99}},
  \bibinfo{pages}{120405} (\bibinfo{year}{2007}).

\bibitem[{\citenamefont{Weld et~al.}(2009)\citenamefont{Weld, Medley, Miyake,
  Hucul, Pritchard, and Ketterle}}]{Weld09}
\bibinfo{author}{\bibfnamefont{D.~M.} \bibnamefont{Weld}},
  \bibinfo{author}{\bibfnamefont{P.}~\bibnamefont{Medley}},
  \bibinfo{author}{\bibfnamefont{H.}~\bibnamefont{Miyake}},
  \bibinfo{author}{\bibfnamefont{D.}~\bibnamefont{Hucul}},
  \bibinfo{author}{\bibfnamefont{D.~E.} \bibnamefont{Pritchard}},
  \bibnamefont{and} \bibinfo{author}{\bibfnamefont{W.}~\bibnamefont{Ketterle}},
  \bibinfo{journal}{Phys. Rev. Lett.} \textbf{\bibinfo{volume}{103}},
  \bibinfo{pages}{245301} (\bibinfo{year}{2009}).

\end{thebibliography}

\end{document}